%% file: main.tex
\documentclass[sigconf]{acmart}

\AtBeginDocument{%
  \providecommand\BibTeX{{%
    \normalfont B\kern-0.5em{\scshape i\kern-0.25em b}\kern-0.8em\TeX}}}

\copyrightyear{2021} 
\acmYear{2021} 
\setcopyright{acmcopyright}\acmConference[CHI '21]{CHI Conference on Human Factors in Computing Systems}{May 8--13, 2021}{Yokohama, Japan}
\acmBooktitle{CHI Conference on Human Factors in Computing Systems (CHI '21), May 8--13, 2021, Yokohama, Japan}
\acmPrice{15.00}

\usepackage{dirtytalk}
\usepackage[utf8]{inputenc}
\usepackage{comment}
\usepackage[english]{babel}
\usepackage{textcomp}
\usepackage{adjustbox}
\usepackage{indentfirst}
\usepackage{mathtools}
\usepackage[usestackEOL]{stackengine}
\usepackage{amsmath}
\usepackage{ltxtable}
\usepackage{algorithm}
\usepackage{algpseudocode}
\usepackage{textcomp}
\usepackage{mathpazo}
\usepackage{tgpagella}
\usepackage{color,soul}
\usepackage{tcolorbox}
\usepackage{color, colortbl} 
\usepackage{flushend}

\usepackage{graphicx,subcaption,chngpage,calc}

\definecolor{lightgray}{gray}{0.9}
\definecolor{DarkGreen}{RGB}{0,176,80} 
\definecolor{DarkGray}{RGB}{208,206,206} 

 \usepackage[math]{cellspace}
  \usepackage{ltablex}
\usepackage{xtab, lipsum, array}
\usepackage{caption}
\usepackage{ltablex}
\usepackage{multirow}
\usepackage{tabularx,ragged2e,booktabs,caption}
\newcolumntype{L}{>{\raggedright\arraybackslash}X}

\stackMath

\DeclareMathSizes{12}{17.28}{9}{7}
\usepackage{color}
\usepackage{graphicx}
\usepackage{array}
\usepackage{multirow, makecell}
  \usepackage{ltablex}

\usepackage{url}
\urldef{\mailsa}\path|{alfred.hofmann, ursula.barth, ingrid.haas, frank.holzwarth,|
\urldef{\mailsb}\path|anna.kramer, leonie.kunz, christine.reiss, nicole.sator,|
\urldef{\mailsc}\path|erika.siebert-cole, peter.strasser, lncs}@springer.com|    
\usepackage{pgf-pie}
\usepackage{tabularx}
\usepackage{pgfplots}
\pgfplotsset{width=7cm,compat=1.8,tick label style={font=\small}}
\usepackage{pgfplotstable}
\usepackage{sfmath}
\usepackage{array, booktabs, makecell}
\usepackage{color}
\usepackage{csquotes}
\usepackage{url}
\usepackage{comment}
\usepackage{tikz}
\usepackage{balance}
\usepackage{listings}
\usepackage{breqn}
\usepackage{booktabs} 
\usepackage{soul} 
\usetikzlibrary{fit} 
\usetikzlibrary{positioning}
\usetikzlibrary{arrows}
\usetikzlibrary{shapes.multipart}
\usepackage[normalem]{ulem}
\usepackage{multirow} 
\usepackage{multicol}
\usepackage{varwidth} 
\newcommand{\eman}[1]{\textcolor{violet}{{\it [Eman: #1]}}}

\newcommand{\wajdi}[1]{\textcolor{blue}{{\it [Wajdi: #1]}}}

\definecolor{Gray}{gray}{0.80} 

\usepackage{mathtools}
\usepackage{amsmath}

\begin{document}

\title{Finding the Needle in a Haystack: On the Automatic Identification of Accessibility User Reviews}


\author{Eman Abdullah AlOmar}
\affiliation{%
  \institution{Rochester Institute of Technology}
  \city{Rochester}
  \state{New York}
  \country{USA}}
\email{eman.alomar@mail.rit.edu}

\author{Wajdi Aljedaani}
\affiliation{%
  \institution{University of North Texas}
  \city{Denton}
  \state{Texas}
  \country{USA}}
\email{wajdialjedaani@my.unt.edu}

\author{Murtaza Tamjeed}
\affiliation{%
  \institution{Rochester Institute of Technology}
  \city{Rochester}
  \state{New York}
  \country{USA}}
\email{mt1256@rit.edu}


\author{Mohamed Wiem Mkaouer}
\affiliation{%
 \institution{Rochester Institute of Technology}
  \city{Rochester}
  \state{New York}
  \country{USA}}
\email{mwmvse@rit.edu}

\author{Yasmine N. Elglaly}
\affiliation{%
  \institution{Western Washington University}
  \city{Bellingham}
  \state{Washington}
  \country{USA}}
\email{elglaly@wwu.edu}


\renewcommand{\shortauthors}{AlOmar et al.}

\begin{abstract}
In recent years, mobile accessibility has become an important trend with the goal of allowing all users the possibility of using any app without many limitations. User reviews include insights that are useful for app evolution. However, with the increase in the amount of received reviews, manually analyzing them is tedious and time-consuming, especially when searching for accessibility reviews. The goal of this paper is to support the automated identification of accessibility in user reviews, to help technology professionals in prioritizing their handling, and thus, creating more inclusive apps. Particularly, we design a model that takes as input accessibility user reviews, learns their keyword-based features, in order to make a binary decision, for a given review, on whether it is about accessibility or not. The model is evaluated using a total of 5,326 mobile app reviews. The findings show that (1) our model can accurately identify accessibility reviews, outperforming two baselines, namely keyword-based detector and a random classifier; (2) our model achieves an accuracy of 85\% with relatively small training dataset; however, the accuracy improves as we increase the size of the training dataset. 
\end{abstract}




\maketitle

\section{Introduction}
\label{ch:Introduction}



Many mobile applications (apps) have poor accessibility which makes it difficult for people with disabilities to use such apps \cite{ yan_current_2019, alshayban2020accessibility, accessandroid, blind_mobile}. Researchers presented several methods, tools, frameworks, and guidelines to support developers in creating accessible mobile applications \cite{ accessmobdesign, chiti2012accessibility, epidemology_framework, ballantyne2018study, eler_automated_2018, tigwell2017ace}. However, many software developers and designers still do not incorporate accessibility into their software development process due to lack of awareness or lack of resources, e.g., budget and time, \cite{ patel2020software,devaccess_services, putnam2012professionals}. In this paper, we present a method that can help software developers to quickly become aware of specific accessibility problems with their apps that the users encountered. Our method is based on automatically identifying app reviews that users write on app stores, e.g., App Store\footnote{https://www.apple.com/ios/app-store/}, Google Play\footnote{https://play.google.com/store} and Amazon Appstore\footnote{https://www.amazon.com/mobile-apps/b?ie=UTF8\&node=2350149011}, where these reviews express an accessibility-related feedback. 

Analyzing app reviews was used by technology professionals to identify issues with their mobile apps \cite{maalej2009users, ciurumelea_analyzing_2017,li_mobile_2018}. 
However, accessibility in user reviews is rarely studied especially for mobile applications \cite{eler_android_2019}. 
Identifying accessibility-related reviews is currently done using two main methods: manual identification and automatic detection \cite{eler_android_2019}. The manual identification approach is time consuming especially with the vast number of reviews that users upload to the app stores, and so it becomes impractical. The automated detection method employs a string-matching technique as a predefined set of keywords are searched for in the app reviews \cite{eler_android_2019}. These keywords were extracted from the British Broadcasting Corporation (BBC) recommendations for mobile accessibility \cite{noauthor_bbc_nodate}. While this method sounds more practical than the manual one, it has its own drawbacks: the string-matching technique ignores that keywords derived from guidelines do not necessarily match the words expressed in reviews posted by users. This mismatch includes but not limited to situations when the keywords are incorrectly spelled by users. 
More importantly, the presence of certain keywords in a review does not necessarily mean that the review is about accessibility. For example, consider the following reviews from Eler et al. dataset \cite{eler_android_2019}: 

\begin{quote}
        \textit{``This is the closest game to my old 2001 Kyocera 2235's inbuilt game 'Cavern crawler'. Everything is so simple and easy to comprehend but that doesn't mean that it is easy to complete right off of the bat. Going into the sewers almost literally blind (sight and knowledge of goods in inventory) is a great touch too. Keep at it. I'll support you at least in donations.''}
\end{quote}

This review contains a set of keywords that could indicate accessibility (e.g., ``\textit{old}'', ``\textit{blind}'' and ``\textit{sight}'') but it is not an accessibility review. In this review, the word ``\textit{old}'' refers to a device rather than a person. The words ``\textit{blind}'' and ``\textit{sight}'' refer to knowledge of goods in the game rather than describing a player's vision. Therefore, the discovery of accessibility reviews heavily relies on the \textit{context}, and so, simply searching for their existence in the review text is inefficient. Due to the overhead of the manual identification, and the high false-positiveness of the automated detection, these two methods remain impractical for developers to use, and so, accessibility reviews remain hard to identify and to prioritize for correction. To address this challenge, it is critical to design a solution with \textit{learning capabilities}, which can take a set of examples that are known to be accessibility reviews, and another set of examples that are not about accessibility but do contain accessibility-related keywords, and learn how to distinguish between them. Therefore, in this paper, \textbf{\textit{we use supervised learning to formulate the identification of accessibility reviews as a binary classification problem.}} 
This model takes a set of accessibility reviews, obtained by manual inspection, in a previous study \cite{eler_android_2019} as input, we deploy state-of-the-art, machine learning models to \textit{learn} the \textit{features}, i.e., textual patterns that are representative of accessibility reviews. 
In contrast to relying on words derived from guidelines, our solution extracts \textit{features} (i.e., words and patterns) from actual user reviews and learns from them. This is critical because there is a \textit{semantic gap} between the guidelines, formally written on an abstract level, and technology-specific keywords. By features, we refer to a keyword or a set of keywords extracted from accessibility-related reviews that are not only important for classification algorithms, but they can also be useful for developers to understand accessibility-related issues and features in their apps. The patterns can be about an app feature that supports accessibility (e.g., ``\textit{font customization}'', ``\textit{page zooming}'' or ``\textit{speed control}''); about assistive technology (e.g., ``\textit{word prediction}'', ``\textit{text to speech}'' or ``\textit{voice over}'') as well as about disability comments (e.g., ``\textit{low vision}'', ``\textit{handicapped}'', ``\textit{deaf}'' or ``\textit{blind}''). 
Particularly, we addressed the following three research questions in our study:

\begin{description}
\item[\textbf{RQ1:}] \textsl{To what extent machine learning models can accurately distinguish accessibility reviews from non-accessibility reviews?}

To answer this research question, we rely on a manually curated dataset of 2,663 accessibility reviews, which we augment with another 2,663 non-accessibility reviews. Then we perform a comparative study between state-of-the-art binary classification models, to identify the best model that can properly detect accessibility reviews, from non-accessibility reviews.
\end{description}

\begin{description}
\item[\textbf{RQ2:}] \textsl{How effective is our machine learning approach in identifying accessibility reviews?}

Opting for a complex solution, i.e., supervised learning, has its own challenges, as models need to be trained, parameter tuned, and maintained, etc. To justify our choice of such solution, we compare the best performing model, from the previous research question, with two baselines: the string-matching method, and the random classifier. This research question verifies whether a simpler solution can convey competitive results.
\end{description}

\begin{description}
\item[\textbf{RQ3:}] \textsl{What is the size of the training dataset needed for the classification to effectively identify accessibility reviews?}

In this research question, we empirically extract the minimum number of training instances, i.e., accessibility reviews, needed for our best performing model, to achieve its best performance. Such information is useful for practitioners, to estimate the amount of manual work needs to be done (i.e., preparation of training data) to design this solution. 
\end{description}

We performed our experiments using a dataset of 5,326 user reviews, provided by a previous study \cite{eler_android_2019}. Our comparative study has shown that the \textit{Boosted Decision Trees} model (BDTs-model) has the best performance among other 8 state-of-the-art models. Then, we compared our BDTs-model, against two baselines: (1) string-matching algorithm and (2) a random classifier. Our approach provided a significant improvement 
in the identification of accessibility reviews, outperforming the baseline-1 (keyword-based detector) by 1.574 times, and surpassing the baseline-2 (random classifier) by 39.434 times.


The contributions of this paper are: 
\begin{enumerate}
    \item  We present an action research contribution that privileges societal benefit through helping developers automatically detect accessibility-related reviews and filter out irrelevant reviews. We make our model and datasets publicly available \footnote{\url{https://smilevo.github.io/access/}} for researchers to replicate and extend, and for practitioners to use our web service and filter down their user reviews. 
    \item  We show that we need a relatively small dataset (i.e., 1500 reviews) for training to achieve 85\% or higher F1-Measure, outperforming state-of-the-art string-matching methods. However, the F1-measure score improves as we add to the training dataset.
\end{enumerate}

\section{Related Work}
\label{ch:RelatedWork}
 It is crucial that mobile applications be accessible to allow all individuals with different abilities to have fair access and equal opportunities \cite{heron2013open}. Prior studies investigated the accessibility issues raised in Android applications \cite{alshayban2020accessibility, vendome2019can}, and others evaluated the accessibility of various websites \cite{agrawal2019evaluating, dominguez2018website, kimmons2017open, wentz2019documenting}. To the best of our knowledge, there is no study classifies user reviews in Android applications using machine learning.

In this section, we highlight several previous works that profoundly influenced our approach. 
We split the related works into three sections: user review, which briefly highlights the role of user reviews in app evolution; accessibility in user review, focuses particularly on detection of accessibility in user reviews; and classification of text documents, where we focus on current approaches in the classification of text such as user reviews by different taxonomies. 

\subsection{User Reviews}
Many researchers concluded that reviews and ratings posted by users on app store platforms can play an essential role in apps’ evolution since most developers consider users’ reviews when working on a new release \cite{ciurumelea_analyzing_2017, li_mobile_2018, palomba_user_2015, pelloni_becloma_2018}. Maalej et al. \cite{maalej2009users} proposed to consider user-input as first means of requirements elicitation in software development. Similarly, Vu et al. \cite{vu_mining_2015} emphasized on the role of users in software lifecycle by developing an approach to identify useful information from users' review. Moreover, Seyff et al. \cite{seyff_using_2010} suggested continuous requirements elicitation from end-users' feedback using mobile devices.

Considering the fact that user reviews can be a powerful driver to mobile app evolution, we are looking into whether we can effectively detect accessibility reviews from users' feedback. This is important because in a highly competitive market, identifying accessibility issues from users' reviews can help developers improve their apps in order to attract more customers and provide better services to users with different abilities. 

\subsection{Accessibility in User Reviews}
Even though user reviews can be a robust tool to mobile apps evolution, and that even mature apps have many trivial accessibility issues \cite{eler_automated_2018, yan_current_2019}, only 1.24\% of mobile app users report accessibility issues to app stores \cite{eler_android_2019}. In other words, 98.76\% of mobile app users do not post accessibility issues in the form of reviews on app stores. In an effort to find whether mobile app users post accessibility-related issues to app stores, 
Eler et al. \cite{eler_android_2019} investigated 214,053 mobile app reviews using a string-matching approach. They depend on a set of 213 keywords derived from 54 BBC recommendations \cite{noauthor_bbc_nodate} proposed for mobile accessibility. In their work, they inspected 214,053 user reviews to identify reviews pertaining to accessibility. Their approach classified a total of 5,076 reviews as accessibility reviews. However, through a manual inspection later, the researchers found that only 2,663 of the reviews were really about accessibility. We used these 2,663 identified accessibility reviews as one of the two groups in our training set required for a supervised machine learning. We created the second group (i.e., non-accessibility reviews) from their total dataset (i.e., 214,053). So far, this is one of the preliminary studies related to the accessibility in mobile app user reviews. 


\subsection{Classification of Text Documents}
Many studies classify app reviews using different taxonomies \cite{ciurumelea_analyzing_2017, di_sorbo_surf_2017,iacob_retrieving_2013,mcilroy_analyzing_2016,panichella_how_2015,pelloni_becloma_2018}, for various purposes: detection of potential feature requests, bug reports, complaints, and praises, etc. Even though many of them identify reviews related to app usability, there is no explicit mention to accessibility related issues \cite{eler_android_2019}. 



Unlike automatic approaches, classification of text documents using a set of \textit{\textbf{predefined keywords}} has been vastly performed across different domains in software engineering. For instance, Eler et al. \cite{eler_android_2019} relied on 213 keywords to identify accessibility-related reviews. Strogylos and Spinelles \cite{stroggylos_refactoring--does_2007} identified refactoring-related commits using one keyword ``\textit{refactor}''. Similarly, Ratzinger et al. \cite{ratzinger2008relation} used 13 keywords to detect refactoring in commit messages. Later, Murphy-Hill et al. \cite{murphy-hill_how_2012} replicated Ratzinger's work in two open-source software using the 13 keywords Ratzinger used. However, they disproved the previous assumption that commit messages in version history of programs are indicators of refactoring activities. The reasoning behind their findings is that developers do not always report refactoring activities as they might associate refactoring activities with other activities such as adding a feature. AlOmar et al. \cite{AlOmar2019IWoR} have also explored how developers document their refactoring activities in commit messages using a variety of 87 textual patterns (i.e., keywords and phrases). Similarly, we believe users can express accessibility concerns without explicitly using any accessibility keywords from the BBC guidelines as assumed by Eler et al. \cite{eler_android_2019}.

In contrast to the keyword-based approaches, we used an automated machine learning approach since learning approaches outperform the accuracy of the keyword-based approach by at least 1.45 times \cite{alomar2020toward,maldonado_using_2017}. 
On the other hand, a keyword-based identification approach (i.e., relying on an existing set of predefined keywords) could generally miss certain reviews, not only because reviews left by users might not always use those keywords to express an accessibility concern, but also because a single word might not be enough to convey an accessibility message. For example, the review ``\textit{I hope someday we change size of the fonts}''; here the context provides an accessibility concern even though the user is not explicitly using keywords such as ``\textit{disabled}'', ``\textit{blind}'' or ``\textit{low vision}''.

\begin{figure*}[ht]
	\centering
    \includegraphics[width=1.0\textwidth,height=\textheight,keepaspectratio]{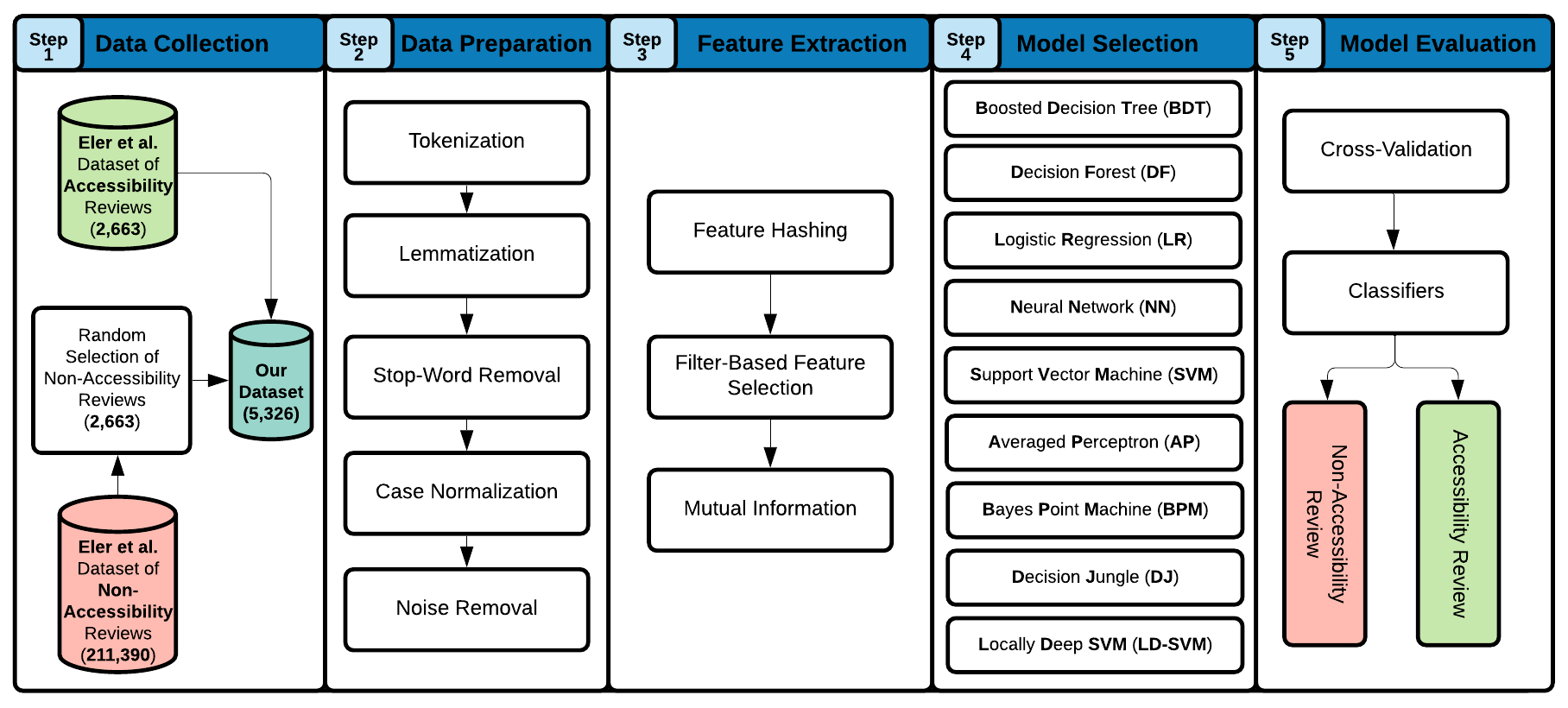}
    \Description[Overall process of accessibility app review]{A five steps process containing data collection, data preparation, feature extraction, model selection and model evolution.}
    \caption{Accessibility app review classification process.} 
    
    \label{fig: Overview}
\end{figure*}

\section{Accessibility App Review Classification} 
\label{ch:Approach}

The main goal of this work is to automatically identify accessibility-related reviews in a large dataset of app reviews. Our approach takes a set of reviews as input and makes a binary decision on whether the review is accessibility pertaining or not, i.e., classifying app reviews (for simplicity we refer to them as \textit{accessibility reviews} and \textit{non-accessibility reviews}). 
To be able to do so, we built a classification model using a corpus of reviews and current classification techniques. We then used the classification model to predict types of new app reviews.  
 Figure \ref{fig: Overview} provides an overview of the process used in the detection of accessibility reviews. Our approach follows five main steps:
\begin{enumerate}
    \item \textbf{Data Collection:} We used a dataset of app reviews along with their ground truth categories previously identified through manual inspection \cite{eler_android_2019} as input for training purposes.
    
    \item \textbf{Data Preparation:} We applied data cleansing and text preprocessing on this set to improve the \textit{reviews text} for the learning algorithms. Some of the text preprocessing procedures we used are namely, tokenizing, lemmatizing, removing stop words, and removing capitalization. 

    
    \item \textbf{Feature Extraction:} We used Feature Hashing \cite{weinberger2009feature}  to extract features (i.e., words) from the preprocessed review text to create a structured feature space. 
    
    \item \textbf{Model Selection and Tuning:} We examined a total of nine classification algorithms to evaluate the performance of the model for prediction. These classifiers were chosen because they are commonly used for classification of text such as app reviews \cite{iacob_retrieving_2013, knauss_detecting_2012}. After training and evaluating the model, we used a testing dataset to challenge the performance of the model. Since the model has already learned from the N-Gram vocabulary and their weights discussed in Section \ref{subsection:DataTransformation} from the training dataset, the classifier output predicted-labels and probability-scores for the testing dataset. Since an app review is a plain text in our case, we follow the approach provided by Kowsari et al. \cite{kowsari_text_2019} that discusses trending techniques and algorithms for text classification, similar to \cite{alomar2020toward,alomar2020we}.
    \item \textbf{Model Evaluation:} We built a training set using the extracted features for the model to learn from.
\end{enumerate}

\subsection{\textbf{Data Collection}}\label{subsection:DataCollection}

\begin{figure}[h]
\centering 
\begin{tikzpicture}
\begin{scope}[scale=0.95]
\pie[rotate = 180,pos ={0,0},text=inside,outside under=30,no number]{20.84/Multimedia\and20.84\%,17.95/Reading\and17.95\%,14.27/Internet\and14.27\%,11.79/Games\and11.79\%,9.01/Writing\and9.01\%,6.61/Phone and SMS\and6.61\%,4.28/Science and Education\and4.28\%,3.30/Time\and3.30\%, 3/Sports and Health\and3\%,2.25/Development\and2.25\%,2.18/Connectivity\and2.18\%,1.92/Navigation\and1.92\%,1.69/Money\and1.69\%,0.90/Security\and0.90\%}
\end{scope}
\end{tikzpicture}
  \Description[Pie chart of accessibility reviews of an app]{reviews divided based on categories multimedia 20.84\%, reading 17.95\%, internet 14.27\%, games 11.79\%, writing 9.01, and other categories.}
  \caption{Distribution of accessibility reviews per app category.}
    \label{fig:reviewsDistPerCategory}
\end{figure}
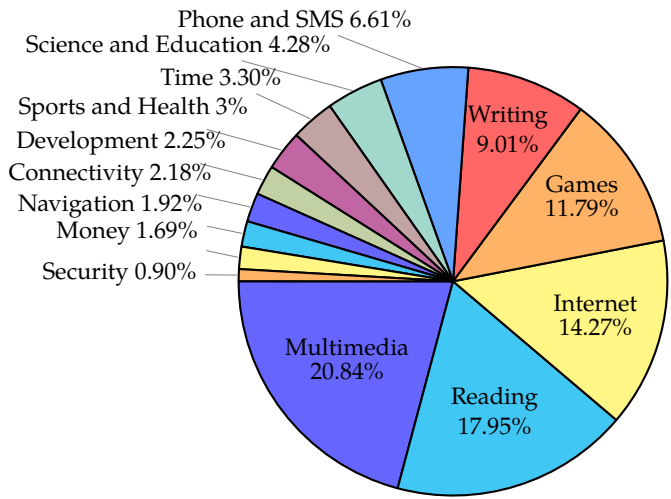

 The dataset, used for this study, and shown in Table \ref{Table:datasetOverview}, is a collection of these 2,663 accessibility reviews, manually validated by Eler et al. \cite{eler_android_2019}. The collected reviews are extracted from across 701 apps, belonging to 15 different categories, as shown in Figure \ref{fig:reviewsDistPerCategory}. This dataset excluded all apps under the Theming and System categories, since they usually do not have any interface associated with them. Eler et al. \cite{eler_android_2019} started with collecting 214,053 reviews, then they performed the string-matching using 213 keywords to filter down reviews and keep only those who potentially may contains information related to accessibility. These keywords are derived from 54 BBC recommendations proposed for mobile accessibility. The string-matching reduced the reviews from 214,053 to 5,076 candidate accessibility reviews. However, the manual inspection of these candidate reviews found that only 2,663 were true positives.
  

\input{Tables/datasetOverview}



In order for us to verify the previous manual labeling of the reviews, we followed the process of Levin et al. \cite{levin_towards_2019} and randomly selected a 9\% sample of reviews, i.e., 243 out of the 2,663 reviews. This quantity roughly equates to a sample size with a confidence level of 95\% and a confidence interval of 6. Then we randomly added another 243 non-accessibility reviews, to end up with a total of 486 reviews. Afterward, one researcher labeled them. The selected data was not exposed to the researcher before. The review process was given a period of 7 days, to avoid fatigue, and the researcher had the opportunity to search online for any keywords they could not understand, during the labeling process. Once the data was labeled, we positioned our labeling against the original labeling of the reviews, from the dataset. We used Cohen's Kappa coefficient \cite{cohen1960coefficient} to evaluate the inter-rater agreement level for the categorical classes. We achieved an agreement level of 0.82. According to Fleiss et al. \cite{fleiss1981measurement}, these agreement values are considered to have an almost \textit{perfect agreement} (i.e., $0.81 – 1.00$).

To prepare training data for the binary classification of app reviews we created two groups of app reviews: (1) reviews indicating accessibility and (2) reviews not related to accessibility. For the accessibility reviews, we used the set of 2,663 reviews previously identified and validated as accessibility reviews through manual inspection by Eler et al. \cite{eler_android_2019}. Since class starvation or an imbalanced training set (i.e., not having equal size of both groups) could decrease the performance of a classification model \cite{levin_boosting_2017,levin_towards_2019}, we need to select an equal number of non-accessibility reviews for the training. 
To efficiently train a classifier, it is important for the negative set to be as \textit{close} as possible to the positive set. Therefore, we chose the negative set to be populated using the discarded reviews of the original authors, during their manual process. These discarded reviews tend to contain some keywords that are relevant to accessibility, but they were found to be conveying another meaning, and that is what we want our model to learn. Since the subset of discarded reviews was 2,413, we randomly selected reviews from the Eler et al. \cite{eler_android_2019} remaining reviews dataset, so that these reviews are also extracted from the same apps, and most likely to contain some keywords that overlap with our true positive set. 

To decide on the number of reviews necessary for training purposes, we reviewed the thresholds used in several text classification studies. The highest number of text documents used in comparable studies \cite{levin_towards_2019,levin_boosting_2017,alomar2020toward} was around 2000 text documents. 
 Since our goal was to provide the model with sufficient reviews that could represent all possible accessibility topics, unlike existing works we chose a total of 5,326 reviews for the model creation and validation. However, we did evaluate our model with different sizes of training sets to understand the size of the training set that yields the best results. We report the results of our evaluations with regard to the testing of different training sizes in Section \ref{ch:ExperimentalResults}.



\subsection{\textbf{Data Preparation}}

Upon completion of the data collection phase, we applied a common approach explained in \cite{kowsari_text_2019} for text preprocessing, similar to \cite{alomar2020toward,alomar2020we}. For a model to classify text documents correctly, the text needs to be cleaned and preprocessed. To preprocess the app reviews text, we used natural language processing techniques, built-in the Microsoft Azure \cite{noauthor_azure_nodate}, such as tokenizing, lemmatizing, removing stop words, and removing capitalization.

\textbf{Tokenization:} is the process of splitting natural text data into tokens, or meaningful elements, that contain no white space. We tokenized app reviews by breaking them into their constituent set of words.

\textbf{Lemmatization:} is the process of getting the basic form of a word by either removing the suffix of a word or replacing the suffix of a word with a different one. It is also the process of reducing the number of unique occurrences of similar words. We used this preprocessing technique to represent words in their canonical form in order to reduce the number of unique occurrences of similar text tokens. 

\textbf{Stop-Word Removal:} We removed words such as (\textit{is, am, are, if, for, the}, etc.) that do not play any good role in classification.  

\textbf{Case Normalization:}
Since we wanted the same words with different font cases (e.g., ``Accessibility'' and ``accessibility'') to be treated as the same word, we converted original review texts to lower case. This type of text cleansing helps us avoid having repeated features differing only in the letter case. We realize that in some cases a user can identify themselves as `Deaf' with uppercase `D' to express their cultural identity in their review which is different from `deaf'. However, as our classifier is a binary classifier that only distinguishes accessibility reviews from the rest, the words `Deaf' and `deaf' will yield the same classification result. Hence, case normalization in this context is safe and will not overrule users' expressions.     

\textbf{Noise Removal:}
We removed any noise that could deteriorate classification performance and confuse the model when learning. Examples of the noise we removed include removing special characters, numbers, symbols, email addresses and URLs.

\subsection{\textbf{Feature Extraction}}\label{subsection:DataTransformation}
\begin{figure*}[ht]
	\centering
    \includegraphics[width=1.0\textwidth,height=\textheight,keepaspectratio]{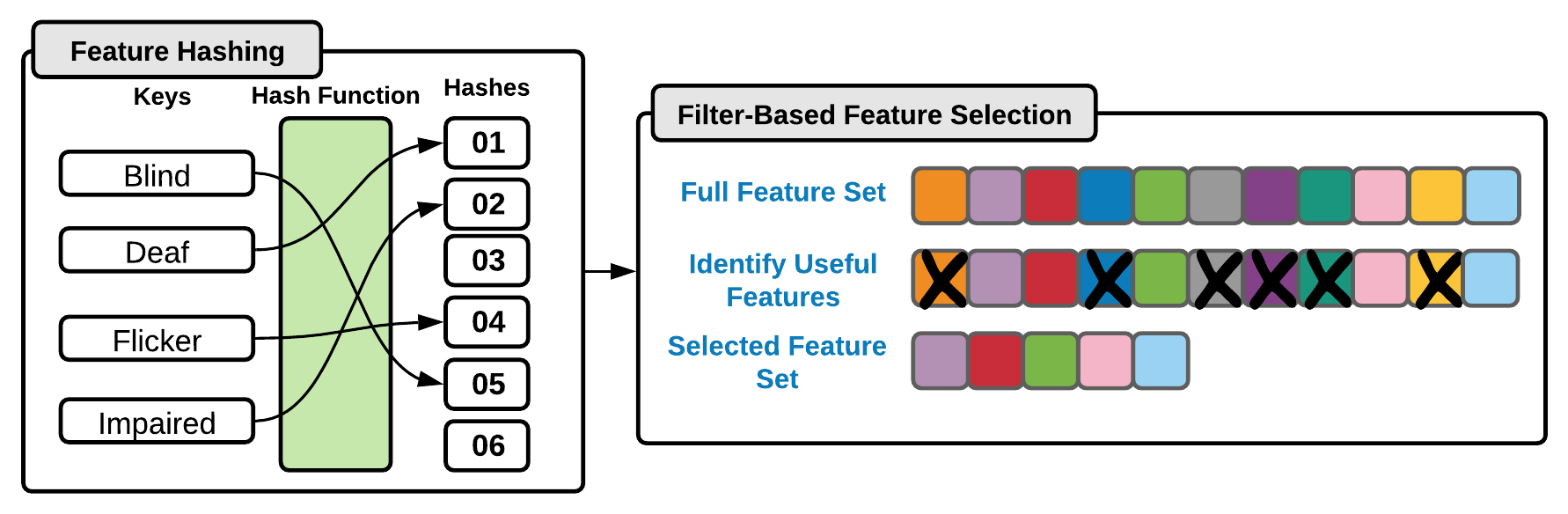}
    \Description[overview approach of feature extraction]{the process split into two parts, feature hashing which hashing the words to vectors, and filter-based feature selection where the useful features are identified}
    \caption{An example of feature hashing and feature selection process in feature extraction stage. 
    } 
    \label{fig: FeatureHashing}
\end{figure*}
After cleansing and preprocessing the reviews text, we extracted features from the preprocessed text that matter the most in distinguishing between the two classes in classification. Particularly, we used the \textit{\textbf{Feature Hashing}} technique for feature extraction. 
Feature Hashing is a technique that operates on high-dimensional text documents used as input in a machine learning model, to map string values directly into encoded features and represent them as integers \cite{shi2009hash,weinberger2009feature}. This technique helps to reduce dimensionality and to make the feature weights lookup more efficient. 
Internally, the Feature Hashing technique creates a dictionary of N-Grams. We used bigrams in our classification since it greatly improves the performance of text classification \cite{tan_use_2002}. Generally, N-Grams have more meaning and semantic than isolated words. For example, the word ``\textit{font}'' does not provide enough information by itself. However, when N-Gram features extracted from reviews, e.g., ``\textit{small font}'', ``\textit{font customization}'', ``\textit{font size}'', etc., the word ``\textit{font}'' can indicate accessibility reviews. We discuss in details the features of our model (i.e., keywords and bigrams) in Section \ref{ch:ExperimentalResults}. We used Mutual Information 
 filter-based feature selection. Mutual Information is a technique that measure how much a variable contributes towards reducing uncertainty about the value of another variable in order to identify features with the greatest predictive power. In fact, this feature set is the training set that the model learns from. In Figure \ref{fig: FeatureHashing}, we illustrate how Feature Hashing applied to the text which was being transformed to a dictionary, as well as the process of the filter-based feature selection.





\subsection{\textbf{Model Selection and Tuning}}

Selecting an appropriate classifier for optimal classification is a challenging task by itself \cite{fernandez2014we}. In this study, we are tackling a two-class classification problem as we are categorizing app reviews into two groups, accessibility and non-accessibility. Because we already have a predefined set of classes, our approach relies on supervised machine learning algorithms to assign each review into one of the two categories. 
We tested nine different classification algorithms as to see which one provides the best results in the context of accessibility and app reviews classification. The tested classifiers are: Logistic Regression (LR), Decision Forest (DF), Boosted Decision Tree (BDT), Neural Network (NN), Support Vector Machine (SVM), Averaged Perceptron (AP), Bayes Point Machine (BPM), Decision Jungle (DJ), and Locally Deep SVM (LD-SVM). We adopted these classifier algorithms because they are commonly utilized in the literature of software-related text classification \cite{lessmann2008benchmarking, goseva2018identification, liaw2002classification, owhadi2019predicting, zhou2017automated,alomar2020toward}. Below is a brief description of each of the classification algorithms used in this study. 
\begin{itemize}
\item \textbf{Logistic Regression (LR)}\cite{andrew2007scalable} is a linear classifiers that predicts the probability of an outcome by fitting data to a logistic function. 
\item \textbf{Decision Forest (DF)}\cite{prinzie2008random}: is a tree-based learner that builds many classification trees. A specific classification is associated with each tree produces. To classify a new object, DF chooses the classification that has the most votes over all other trees.
\item \textbf{Boosting Decision Tree (BDT)}\cite{friedman2001greedy}: is an ensemble learning method in which the second tree corrects for the errors of the first tree, the third tree corrects for the errors of the first and second trees, and so forth. Predictions are based on the entire ensemble of trees together that makes the prediction.
\item \textbf{Neural Network (NN)}\cite{hansen1990neural}: is a set of interconnected layers. The inputs are the first layer that are connected to an output layer by an acyclic graph. 
\item \textbf{Support Vector Machine (SVM)} \cite{wu2008top}: is a learner that constructs hyperplane(s) in n-dimensional space. 
\item \textbf{Averaged Perceptron (AP)}\cite{collins2002discriminative} is a simple version of Neural Network. The inputs are classified into several outputs based on a linear function, and then combined with a set of weights that are derived from the feature vector. 
\item \textbf{Bayes Point Machine (BPM)}\cite{herbrich2001bayes}: is an algorithm that uses a Bayesian approach to linear classification called the ``Bayes Point Machine''. This algorithm approximates the optimal Bayesian average by choosing one ``average'' classifier, the Bayes Point.
\item \textbf{Decision Jungle (DJ)}\cite{shotton2013decision}: is a recent extension to decision forests. It consists of an ensemble of decision directed acyclic graphs (DAGs).
\item \textbf{Locally Deep SVM (LD-SVM)}\cite{jose2013local}: is a classifier that has been developed for an effencient non-linear SVM prediction. 
\end{itemize}

We compared all the nine classifiers based on their common statistical measures such as precision, recall, accuracy, and F1-measure. These experiments were performed on the Azure ML platform because it provides a built-in web service once the classification model is deployed. We report the results of our classifier comparison and evaluation in Section \ref{ch:ExperimentalResults}. 

We use grid search cross validation \cite{gridS}, a tuning method that performs exhaustive search over specified parameter values for an estimator, for tuning of our selected ML models. In order to facilitate the replication of our results, we provide the selected main parameters for ML techniques as shown in Table \ref{tab:HyperparameterML}.

\input{Tables/HyperparameterML}

\subsection{\textbf{Model Evaluation}}

We assess the performance of our selected models based on the following four measurement aspects: 
\begin{itemize}
  \item \textbf{Precision} = $\frac{tp}{tp+fp}$: is a statistic that calculates the accurate number of correct predictions out of all the input sample.
   \item \textbf{Recall} = $\frac{tp}{tp+fn}$: is a statistic that calculates the accurate number of positive predictions that was actually observed in the actual class. 
  \item \textbf{Accuracy} =  $\frac{TP + TN}{TP + TN +FP + FN}$: is a statistic that calculates the accurate number of
  \item \textbf{F1-measure} = $\frac{2 \cdot P\cdot R}{P+ R}$: is a a statistic that calculates the accuracy from the precision and recall.
\end{itemize}

Here TP denotes True Positive, TN denotes True Negative, FP denotes False Positive, and FN denotes False Negative. These metrics participation in measurement for a classifier’s output.
\begin{itemize}
   \item \textbf{True Positive (TP):} This parameter determines the predictions labeled correctly by the classifier as positive.
   \item \textbf{True Negative (TN):} This parameter determines the correct number of negative predictions.
  \item \textbf{False Positive (FP):} This parameter determines the number of instances (negatives) that were presumed as positive instances by the classifier by mistake.
  \item \textbf{False Negative (FN):} This parameter determines the number of positive instances that were falsely assumed to be as negative instances by the classifier. 
\end{itemize} 

\textbf{Cross-Validation.} We applied a 10-fold cross-validation technique to evaluate the variability and reliability of our models. For each model, we split our dataset into 10 folds containing the equal size of app reviews. Then, we performed 10 evaluations with various testing datasets wherein each evaluation 9 folds were used as a training dataset and the other fold was used as a testing dataset. Put differently, unlike other approach that is dependent on just one train-test split, when evaluating our model using 10-fold cross-validation, we train on multiple train-test splits in which one fold is left as a holdout data set, so it is unseen during the training. This approach is considered the preferred method as it gives us a better indication of how well our model performs on unseen data. We aggregated the results of the 10 evaluations and reported the average performance tested with multiple models.

\section{Experimental Results and Evaluation}\label{ch:ExperimentalResults}

In this section, we review the results of our experiments to evaluate the performance of our approach. For evaluating various accessibility classification models, we used standard statistical measures (\textit{Precision}, \textit{Recall}, \textit{Accuracy}, \textit{F1-measure}). Using the evaluation results, we provide answers to our research questions. 

 
\label{section: RQ1}
\textit{\textbf{RQ1. To what extent machine learning models can accurately distinguish accessibility reviews from non-accessibility reviews?}}


We conducted an experiment to determine if the automatic classification of user reviews using machine learning techniques can be performed with high accuracy. 
We wanted to understand the opportunities and limitations of the machine learning technique in automatically detecting accessibility reviews. 

\begin{figure*}[ht]
	\centering
    \includegraphics[width=1.0\textwidth,height=\textheight,keepaspectratio]{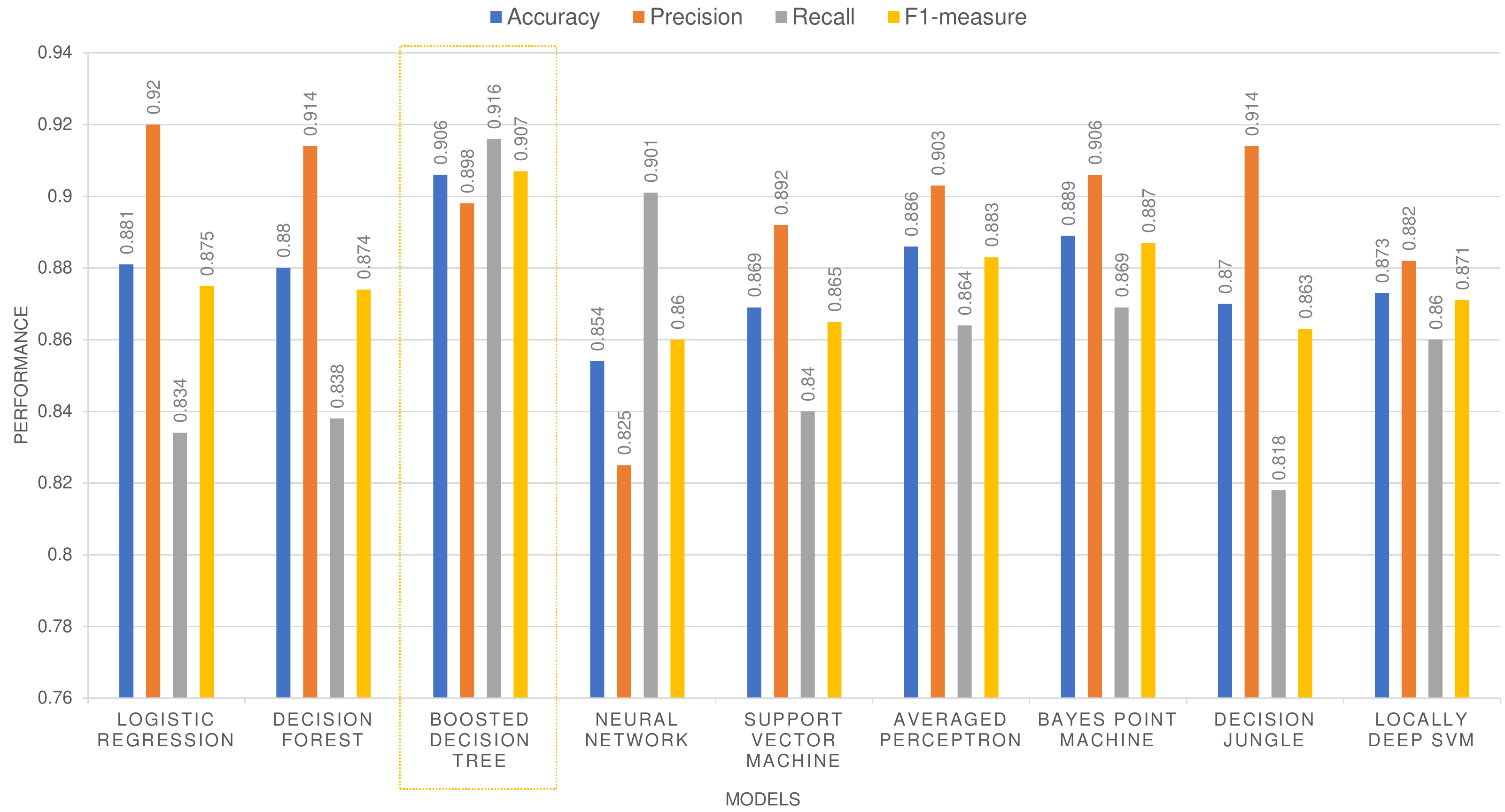}
    \Description[bar chart presenting the evaluation of nine classifiers]{All nine classifiers presented their evaluation parameters, which consist of precision, recall, accuracy and F1-measure}
    \caption{Comparison between binary classifiers, in terms of precision, recall, accuracy, and F1-measure.}
    \label{fig:F-1}
\end{figure*}

We compared the nine classification algorithms tested in this study with respect to precision, recall, accuracy and F1-measure and reported the results as shown in Figure \ref{fig:F-1}. The accuracy and F1-measure of the Boosted Decision Trees model (BDTs-model) is clearly higher than its competitors for the classification of accessibility reviews. The BDTs-model with the accuracy of 90.6\% and F1-measure of 90.7\%, outperformed other classification algorithms. Figure \ref{fig:F-1} also shows that the Bayes Point Machine (BPM) and Averaged Perceptron (AP) with F1-measure of 88.7\% and 88.3\% respectively, yielded higher predictive power after the Boosted Decision Trees. 

The fact that BDTs-model achieved top performance rate can be explained by the fact that a boosted decision tree aggregates several learnings since it is an ensemble learning method. In the ensemble method, the errors of the first tree are fixed by the second tree, and the errors of the second tree are fixed by the third, and so on. In this method, the entire ensemble trees together form the prediction. 

To further understand how these models distilled the text of the reviews into features, we extract keywords that were trending in our dataset, that we enumerate in Table \ref{Table:KeywordsHavingReviews}. It is important to note that the majority of these keywords were identified by the BBC recommendations for mobile accessibility, however, not all of these keywords were found to be useful for our best performing classifier, i.e., BDTs-model. In Table \ref{Table:KeywordsHavingReviews}, we report in bold, the features that were influential in increasing the accuracy of the trained Boosted Decision Trees. Such finding does not necessarily deny the relevance of the remaining keywords in describing accessibility related issues, but the fact that they were not selected, indicates their existence in non-accessibility related reviews. Keywords such as ``\textit{dark mode}'' or ``\textit{mute}'', while being used in the BBC guidelines, are also known to be used in general usability contexts. For example, the keyword ``\textit{mute}'' tends to be frequently used in reviews related to media and video players, where sound is one of the main features of the app.  

Further, on a more qualitative sense, we examine the set of frequently occurring bigrams for the keywords (reported in Table \ref{Table:KeywordsHavingReviews}) that are strongly correlated to the accessibility review . Bigram corresponds to a sequence of two adjacent words in a sentence to help better understanding the context for the given terms. By analyzing the natural language in the accessibility review, we obtain more specific accessibility review-related terminology. Table \ref{Table:bigram} presents the frequently occurring bigrams in the review. Looking at these terms, we see that developers are either commenting on the features of the apps (e.g., ``easily accessibile'', ``good text reflow'', ``great for visually impaired''), or they are discussing accessibility issues with their products pointing out that the apps need to be improved (e.g., ``terribly hard to see'', ``no visual cue'', ``cant read''). 


 The findings, illustrated in Tables \ref{Table:KeywordsHavingReviews} and \ref{Table:bigram} indicate a potential variation of how users typically state their accessibility needs. While it seems intuitive, there are no studies that focused on extracting such information in a structured manner to facilitate the identification of such accessibility problems by the app maintainers.
 
  Although a high classification performance of our BDTs-model has been demonstrated in Figure \ref{fig:F-1}, there are some limitations that lead BDTs-model to output some misclassified reviews as illustrated in Table \ref{Table:misclassification}. According to our thorough analysis, we notice that the misclassification of our model can be related to:
 \begin{itemize}
     \item False positive instances caused by the format of reporting user perspective of the apps. The examples in the table show that different expression about the apps like ``simple'' or ``headache'' can be confusing to the classifier and hence it misclassified these reviews.  
     \item False negative instances caused by the format of reporting a specific feature of the apps. As shown in the table, the users commented on a specific feature such as ``functioning reader'' and ``caller ID''. The BDTs-model will wrongly classify it because these could be seen as an accessibility-related features. 
 \end{itemize}
 It is worth noting that the above misclassifications do not have a large influence on the overall performance of the BDTs model. Only a small number of reviews are wrongly classified by our model. 

\begin{tcolorbox}
\textit{Summary}. 
The Boosted Decision Trees model, 
with an accuracy of 90.6\% and an F1-measure of 90.7\%, is the best performing model in the binary classification of accessibility reviews.  
\end{tcolorbox}

\input{Tables/KeywordsHavingReviews}
 \input{Tables/Bigrams}

\input{Tables/Misclassification}

\textbf{\textit{RQ2. How effective is our machine learning approach in identifying accessibility reviews?}}



The main goal of this study is to propose an automatic approach for identification of accessibility reviews that can effectively outperform current state-of-the-art baselines: Keyword-based (i.e., also called pattern-based or string-matching) \cite{eler_android_2019} and Random classifier \cite{maldonado_using_2017}. Existing studies that have applied machine learning techniques in similar contexts (i.e., text classification) usually evaluate their approach using different classifiers. To compare their approach against others, they consider the keyword-based approach. To our knowledge, the only study that considers additional approach (i.e., random classifier) is the study by Da Silva et al. \cite{maldonado_using_2017}. Thus, we consider keyword-based and random classifier to compare against our approach. 
Answering this question is important to understand if the detection of accessibility reviews is a learning problem. We hypothesize that learning algorithms can outperform string-matching algorithms. 
To examine if the hypothesis holds true, we chose to investigate the following two baselines, and compared them with our BDTs-model.


\textbf{Baseline 1. Keyword-based Approach.}
The keyword-based (string-matching) approach for identifying accessibility reviews is suggested by Eler et al. \cite{eler_android_2019}. In their work, they inspected 214,053 user reviews to identify the reviews pertaining to accessibility. Their string-matching approach classified a total of 5,076 reviews as accessibility reviews. However, manual verification of the 5,076 reviews later found that only 2,663 of the reviews were correctly identified \cite{eler_android_2019}. 

To calculate statistical metrics for baseline 1, we used a set of 5,326 reviews (cf., set of 2,663 accessibility reviews, from Table~\ref{Table:datasetOverview}, and another 2,663 non-accessibility reviews, selected from the same apps). Then, we manually inspected 
 these reviews to determine true positives (\textit{TP}), true negatives (\textit{TN}), false positives (\textit{FP}), and false negatives (\textit{FN}). 
True positives are when the keyword-based approach correctly detected accessibility reviews, and true negatives are when non-accessibility reviews are correctly identified. 
False positives are the reviews identified as accessibility reviews while they are not; and false negatives are the reviews identified as non-accessibility reviews while they are accessibility reviews. Since we already had the reviews labelled, 
 we were able to count \textit{TP}, \textit{TN}, \textit{FP} and \textit{FN}.



\textbf{Baseline 2. Random Classifier.}
Similar to Da Maldonado et al. \cite{maldonado_using_2017}, we consider Random classifier as one of the baselines to compare our approach to. 
The precision of the random classifier technique  
is calculated by dividing the number of accessibility reviews by the total number of user reviews (i.e., $ \frac{2663}{214053} = 0.012 $). When it comes to recall, there is only 50\% probability for a review to be classified as an accessibility review since there are two possible classifications available. Finally, the F1-measure of baseline 2 is calculated as $ 2* \frac{0.012*0.5}{0.012+0.5} = 0.023 $. 

Using the values of \textit{TP}, \textit{TN}, \textit{FP} and \textit{FN}, we calculated the Precision, Recall, and F1-measure, for both baselines.
Table \ref{Table:Approaches comparision} shows the standard statistical measures of the three approaches, also the performance improvements 
achieved by our BDTs-model 
compared to the other two methods.

As can be seen from Table \ref{Table:Approaches comparision}, F1-measure obtained by the machine learning approach is much higher than the other methods. 
 F1-measure achieved by the machine learning approach is 0.90, while F1-measure values using keywords and random classifier are 0.576 and 0.023 respectively. 
Table \ref{Table:Approaches comparision} shows that our approach outperforms the keyword-based approach by 1.574 times and the random classifier by 39.434 times when identifying accessibility reviews. To better understand the performance of the string-matching method, we have extracted examples reviews that were wrongly classified, as accessibility: 

\begin{quote}
        \textit{Review 1. ``Good to have your files easily accessible. Would like integration of caldav/ carddav''} 
\end{quote}

\begin{quote}
        \textit{Review 2. ``Very useful application. 
Gmail users must go for it blind eyes''} 
\end{quote}

The existence of keywords such as ``\textit{accessible}'' and ``\textit{blind eyes}'', are string-matched to the keywords considered as accessibility by the guidelines, and so, the keyword-based approach will flag their corresponding reviews as accessibility. However, the first review (i.e., Review 1) refers to the new feature that allows user files to be accessible more efficiently and requests the integration of a protocol for the synchronization of calendars. Similarly, the second review (i.e., Review 2), is praising an app that synchronizes Gmail calendar with Outlook calendar, and the user's expression of "going with blind eyes", refers to their satisfaction, and not to what would be considered by the string-matching method as an accessibility issue. 


To determine the different cases of when the keyword-based approach fails, we evaluated 592 reviews, a statistically significant sample with a confidence level of 99\% and a confidence interval of 5\%. By analyzing the selected reviews, we identified the following reasons behind the failure cases of the string-matching approach:
\begin{itemize}
    \item \textbf{Keyword Misspelling.} This category depicts the case when accessibility aspects of the mobile application are addressed by the users using misspelled keywords. This case can be illustrated in the following example: \textit{`` Font size of lowercase letters is sooosmall! How to change it? It should be like on google keybord when you change capital/lowercase mode - lowercase letters have almost the same size as capital. It's much easier for your eyes!''}. The keyword matching approach can miss any word with a typo or with inproper spacing, such as ``\textit{keybord}'' or ``\textit{sooosmall}''. 
    Misspellings are frequent in app user reviews, since mobile writing is known to be more prone to typos.  
    \item \textbf{Keyword Variation.} This category shows the case in which users use different part-of-speech (POS) of the accessibility-related keywords reported in Table \ref{Table:KeywordsHavingReviews}. As shown in the following review: \textit{``very accessible as a blind user thank you''}, the user used the adjective form (``accessible'') of the word accessibility. 
    \item \textbf{Expression Variation.} This category represents cases in which users use different expressions of the keywords listed in Table \ref{Table:KeywordsHavingReviews} to address accessibility aspects of the apps. This case is best illustrated in the following accessibility review: \textit{``still getting responses from the wrong people and noticed that when in night mode with pure black background - when you try to delete a message the yes option is completely black so impossible to see''}. As can be seen, the expression ``impossible to see'' is used instead of the keyword ``cannot see'' to represent the user perspective on the problem. 
\end{itemize}
\input{Tables/DifferentApproachComparison}

\begin{tcolorbox}
\textit{Summary}. 
The Boosted Decision Trees model outperforms the current state-of-the-art approaches in the classification of accessibility reviews. We obtained an F1-measure score of 90.7\% with an improvement of 1.574x and 39.434x over the keyword-based and random classifier approaches respectively. 
\end{tcolorbox}


\label{section:TrainingSize}
\textbf{\textit{RQ3: What is the size of the training dataset needed for the classification to effectively identify accessibility reviews?}}

So far, we showed that our machine learning approach can accurately identify user reviews that pertain to accessibility. However, 
the performance of a classifier relies on the size of the training data. At the same time, creating a training dataset is a challenging and time-consuming task. Thus, the question is: What is the size of the training dataset needed to effectively classify user reviews? If an approach requires a very large training dataset than it will require a considerable time and effort to be applied to other similar contexts. However, if less training dataset is required to effectively classify accessibility reviews, then our approach can be applied and extended with little efforts. 

To answer this research question, we incrementally added reviews to the training dataset and evaluated the performance of the classification. We began by creating a large training dataset that contains equal size of accessibility reviews and non-accessibility reviews. Then, we used cross validation technique, which is a technique that partitions the original dataset into a training set to train the model, and a test set to evaluate it using number of folds \cite{kohavi1995study}. In this study, we divided the dataset into 10 folds making sure they contain equal size of both classes. Next, we tested our approach using a 10-fold cross-validation technique  
using 9 folds for training and 1 fold for testing. Since we wanted to monitor the performance of our classifier as the training dataset size increased, we incrementally added batches of 100 reviews 
until we used all of our training data (e.g., 5,326 reviews). It is important to note that we considered the equal size of accessibility reviews and non-accessibility reviews with batches incrementally added to the training dataset. We computed the F1-measure value for each iteration (e.g., after adding batches of new reviews to the training set). 
We recorded the number of reviews needed to achieve at least an F1-measure of 80\% to 90\%.

Figure \ref{Chart:TrainingSizeImpact} shows F1-measures calculated when detecting accessibility reviews, while incrementally adding batches of reviews to the training dataset. Our results show that the highest F1-measure (i.e., 0.907) was achieved with 5,326 reviews (our total training dataset) and the lowest F1-measure value (i.e., 0.630) was achieved with 100 reviews. Our results also show that 80 to 90 percent F1-measure is achieved with 400 to 5000 reviews in the training dataset. Such that, we need only 400 reviews to get around 80\% F1-measure and we need at least 1500 reviews to get 85\% or higher, while with 5000 reviews we got around 90\% F1-measure. Finally, we found that the F1-measure score improves as we add to the training dataset.


\input{Charts/TrainingDataSize}

\begin{tcolorbox}
\textit{Summary}. 
We find that we need a relatively smaller training dataset (i.e., 1500 reviews) to get 85\% or higher F1-measure. The F1-measure score improves as we add to the training dataset.    
\end{tcolorbox}


 

\section{Discussion}

We presented a new approach that identifies app reviews with accessibility concerns. We compared our new approach to the current state-of-the-art methods. Based on these findings we discuss implications that can be theory-based and practice-based. Theory-based implications show how this study can further advance the research on accessibility reviews. Practice-based implications show how our model supports our community in building and maintaining accessible mobile apps.

\textbf{Implication 1: App reviews are rich source of information that can be mined to identify specific accessibility problems with the mobile app. }
There are so many accessibility guidelines that developers and designers can find it difficult to test for all of these guidelines. Additionally, adhering to these guidelines does not necessarily guarantee the accessibility of the said app. Also, usability testing with different groups of people with disabilities, e.g., blind or deaf, can be infeasible especially for medium and small-scale companies. One way to discover accessibility problems which prior testing did not reveal is to listen to the users and learn from the reviews they wrote. Our approach can aid technology professionals to quickly spot accessibility problems with their app. 


\textbf{Implication 2: Accessibility as part of mobile apps maintenance and evolution.}
There exist accessibility testing tools and methods that are designed to support the implementation and testing phases of the software. However, there are no tools, to the best of our knowledge, that supports software accessibility in the maintenance phase. With changes made to an app, either for adding a feature or fixing a bug, accessibility can be at risk. Also, with updates made to the phone’s operating system or the installed assistive technology, the accessibility of an app may deteriorate. We call for innovative methods that can support technology professionals in maintaining the accessibility of their app after its release. Our approach in analyzing app reviews offers an opportunity for developers and designers in detecting accessibility pitfalls based on their users’ written feedback. However, with the tremendous number of reviews developers receive on a daily basis, it becomes impractical to manually read through them and identify potential issues related to their new release. Adding our model to the pipeline, will alleviate the manual overhead of looking up accessibility related reviews, and so developers can quickly locate their corresponding issues, and add them to their maintenance pipeline.

\textbf{Implication 3: Understanding users’ language in expressing their accessibility concerns.}
When we compared our BDTs-model to the keyword-based detector, we found that some accessibility reviews did not contain the accessibility keywords that were driven from accessibility guidelines \cite{eler_android_2019}. This indicates that users voice their accessibility feedback using ``user taxonomy'' which may or may not echo the technical and professional terms used in accessibility standards. Further research is needed to understand how users describe mobile accessibility issues. By learning the accessibility ``user taxonomy'', we can improve our BDTs-model, which will lead to enhanced discovery of accessibility reviews. 

\textbf{Implication 4: The interplay between developers and designers, accessibility experts, and users.}
Accessibility experts establish guidelines and design methods in support of creating accessible software. Technology professionals often are not able to digest all these guidelines and often find existing resources lacking. This situation yielded to the existence of software products that are inaccessible to people with disabilities. The effective involvement of people with disabilities in this process can help bridging the communication gap between accessibility experts and developers and designers. By giving users the opportunity to lead the prioritization of accessibility issues based on their usage experience, mobile apps accessibility can be improved in a more meaningful way for people with disabilities. Analyzing app reviews is one way to give users the lead in determining which accessibility issue should be fixed in the next release. Analyzing app reviews can also offer insights to accessibility experts on users’ accessibility needs right from the field, which will be more realistic than results collected from controlled lab studies.
 

\textbf{Implication 5: Direct and immediate apps filtering benefit for end users.} People find online reviews helpful in making purchase decisions \cite{baek2012helpfulness}. Peer comments help users become aware of the limitations of reviewed products \cite{mudambi2010research}. 
Currently, on mobile applications stores, e.g., App Store and Google Play, users can read all reviews, sort them by most helpful or most recent. However, mobile application stores provide no means to filtering reviews based on relevance to specific quality metrics, e.g., accessibility. This lack of filtering pushes users to download the app first and then experience its accessibility, leaving no room for benefiting from peer comments. Sometimes, apps suffer from accessibility regression giving users an unpleasant surprise with an updated app that is less accessible than its former version \cite{sina_tweet}. We call on mobile application stores to take action and allow users to filter reviews based on relevance to accessibility.


\textbf{Implication 6: Pushing the boundaries of Accessibility testing.}
Current accessibility testing strategies are human intensive, and therefore become expensive and impractical, as most developers struggle to find the appropriate testers who can evaluate the compliance of their apps to accessibility guidelines. Existing accessibility scanners are tailored for the web, and they cannot be applied to the mobile environment. In this context, online user reviews, offer a rich source of scenarios, which can be coupled with the app's current version, to create test cases of practically captured anomalies. Relying on this set of reviews, as a shared knowledge, developers can quickly identify potential test cases that they need to perform, in case they are incorporating a given accessibility tool in their app. Furthermore, as the mobile environment is extremely dynamic, recent user reviews can quickly reveal any appearing anomalies in the newer app releases.

\section{Threats to Validity} \label{ch:Threats}
In this section, we identify several threats to the validity of our study. We group the threats to Construct Threats and External Threats to validity.  


\textbf{Construct Threats} relate to the appropriateness of our dataset and accuracy of the previous work \cite{eler_android_2019}. A potential threat is related to creating a training dataset or the manual classification. Developing a training dataset is typically a tedious job, also subject to reader bias. We mitigated this risk by choosing a dataset of accessibility reviews as our training data that were previously identified and validated \cite{eler_android_2019}. Additionally, 
we used all of the identified reviews as training input rather than choosing a sample set of reviews. A total of 2,663 reviews were previously identified as accessibility reviews from 214,053 app reviews through manual inspections and validations. 

Another potential threat relates to the keywords used for the identification of accessibility reviews through a string-matching approach. The string-matching approach relied on 213 keywords derived from 54 accessibility recommendations by BBC. The keywords and phrases users use in their reviews do not necessarily match the keywords available in the guidelines and recommendations. This mismatch includes but not limited to situations when keywords would be spelled incorrectly by reviewers. A related concern is whether the set of keywords is inclusive of all possible keywords that users use to express their accessibility concerns. To mitigate this threat, we used keywords defined by \cite{eler_android_2019} in which the authors adopted variants for these keywords to ensure they would not miss any relevant review during their manual validation. This raised our confidence to use the dataset that has these keywords as a representative sample of accessibility reviews.  


\textbf{External Threats} relate to the generalizability of our findings for this evaluation. We evaluated and tested our findings on a dataset collected by previous researchers \cite{eler_android_2019}. The dataset was collected only from Android open-source applications. Therefore, the dataset did not represent the entire mobile apps on the App stores such as Apple store applications. Also, we only study mobile application reviews of open-source applications. Our results may not generalize to commercially developed projects or to other reviews that are written in other languages than English. 

\section{Conclusion} \label{ch:Conclusion}


This study presents an approach that automates the classification of app reviews as accessibility-related or not so developers can easily detect accessibility issues with their products and improve them to more accessible and inclusive apps utilizing the users' input. As Hayes pointed out:\say{In Action Research, the goal is ultimately to create sustainable change. That is to say, once the research facilitators leave, the community partners should be able to maintain the positive changes that have been made.} \cite{hayes2011relationship}. Our goal is to create a sustainable change, by including a model in developer’s software maintenance pipeline, and raising awareness of existing errors that hinders the accessibility of mobile apps, which is a pressing need \cite{patel2020software}.

As we develop our model, we conducted an evaluation of nine different classifiers using an existing dataset of manually validated accessibility reviews. Our evaluation shows that the Boosted Decision Tree classifier offers higher accuracy than the other approaches in the classification of app reviews. Additionally, we compared our approach with two baselines, namely a keyword-based approach, and a random classifier. The results indicate that our approach outperforms the two state-of-the-art approaches with the F1-measure of 90.7\%. Finally, we conduct an experiment to evaluate the impact of training data sizes on our classifier’s accuracy. Our evaluation shows that we need a relatively smaller dataset (i.e., 1500 reviews) for training to get 85\% or higher F1-measure. However, the F1-measure score improves as we add to the training dataset.


As our results show, having an adequately large training size is important for high accuracy in prediction. Given the millions of app reviews available on the app store platforms, the training process can be cumbersome and laborious. Additionally, it is necessary to obtain labels from multiple Subject Mater Experts (SMEs) to make the training dataset more reliable. In order to further reduce the efforts needed by developers and SMEs in creating a training data, we are planning to explore \textit{Active Learning} \cite{settles_active_2012, settles_analysis_2008}, a well-known machine learning paradigm for classification. We also plan to perform a multi-class classification on the accessibility reviews --- dividing them into categories such as readability of text, audio, video, UI, gestures etc. 

\begin{acks}
The authors would like to thank William Catzin for his help in this work. This material is based upon work supported by the National Science Foundation, USA, under Grant No. 1757680.
\end{acks}

\balance
\bibliographystyle{ACM-Reference-Format}
\bibliography{sample-base}

\end{document}

%% file: Tables/datasetOverview.tex
\begin{table}[ht]
\centering
\caption{Statistics of the dataset.}
\label{Table:datasetOverview}

\begin{tabular} {|c|c|}\hline

{\cellcolor{lightgray}\textbf{Number of Apps}}                    & 701   \\ \hline
{\cellcolor{lightgray}\textbf{App Categories}}          & 15    \\ \hline
{\cellcolor{lightgray}\textbf{All Reviews}}             & 214,053 \\ \hline
{\cellcolor{lightgray}\textbf{Accessibility Reviews}}   & 2,663 \\ \hline

\end{tabular}
\end{table}

%% file: Tables/HyperparameterML.tex
\begin{table*}[htbp]
\caption{Summary of the hyperparameter in machine learning algorithm.} 
\label{tab:HyperparameterML}
\resizebox{\textwidth}{!}{%
\begin{tabular} {llll}
\hline\toprule

\textbf{Classifier} & \textbf{Hyperparameter} & \textbf{Default} & \textbf{Description} \\ \hline    

                     & optimiz_tol             &  1E-07  &  Optimization tolerance \\ 
\textbf{LR}          & 1_weight                &  1         &  L1 regularization weight\\ 
                     & L2_weight               &  1         &  L2 regularization weight \\ 
                     & memory_L_BFGS           &  20        &  Memory size for L-BFGS \\ \hline

                     & n_estimators      & 8     & Number of decision trees   \\ 
\textbf{DF}          & max_depth         & 32    & Maximum depth of the decision trees   \\ 
                     & n_samples_leaf    & 125   & Number of random splits per node    \\ 
                     & min_samples_split & 1     & Minimum number of samples per leaf node   \\ \hline

                     & max_n_leaf          & 20    & Maximum number of leaves per tree    \\
\textbf{BDT}         & min_samples_leaf    & 10    & Minimum number of samples per leaf node \\
                     & learning_rate       & 0.2   & Learning rate  \\
                     & n_tree              & 100   & Number of trees constructed    \\ \hline

                     & n_nodes                & 100   & Number of hidden nodes \\ 
                     & learning_rate          & 0.1   & Learning rate \\ 
\textbf{NN}          & n_learning_rate        & 100   & Number of learning iterations    \\ 
                     & learning_rate_weights  & 0.1   & Initial learning weights diameter   \\ 
                     & momentum               & 0     & Momentum  \\ \hline

                     & n_iter            & 1       & Number of iterations \\ 
\textbf{SVM}         & Lambda            & 0.001   & Lambda \\ \hline


                     & learning_rate     & 1   & Learning rate  \\ 
\textbf{AP}          & m_iter            & 10  & Maximum number of iterations \\ \hline


\textbf{BPM}         & n_training_iter     &  30 & Number of training iterations \\  \hline

                     & n_estimators      & 8     & Number of decision directed acyclic graphs \\ 
\textbf{DJ}          & max_depth         & 32    & Maximum depth of the decision directed acyclic graphs \\ 
                     & max_width         & 128   & Maximum  of the decision directed acyclic graphs    \\ 
                     & n_optimiz         & 2048  & Number of optimization steps per decision directed acyclic graphs layer  \\ \hline


                     &  max_depth         &  3       &  Depth of the tree  \\ 
                     &  lam_weight        &  0.1     &  Lambda weight \\ 
\textbf{LD-SVM}      &  n_theta           &  0.01    &  Lambda Theta  \\
                     &  n_theta_Prime     &  0.01    &  Lambda Theta Prime  \\
                     &  n_sigmoid         &  1       &  Sigmoid sharpness  \\
                     &  n_iter            &  15000   &  Number of iterations  \\

\bottomrule \hline

\end{tabular}

}

\end{table*}

%% file: Tables/KeywordsHavingReviews.tex
\begin{table*}[tbp]
\begin{center}
\caption{List of keywords trending in the 5326 reviews. Keywords in \textbf{bold} are found to be strongly correlated to accessibility reviews by our model. 
}
\label{Table:KeywordsHavingReviews} 
\begin{adjustbox}{width=1.0\textwidth,center}
\begin{tabular}{lllll}

\toprule
\textbf{Keywords} \\
         \midrule 
        (1) dark mode                & (16) adjustable          & (31) \textbf{voice command}    & (46) colour coding          & (61) captcha                      \\
        (2) zoom                     & (17) \textbf{blind}      & (32) \textbf{text-to-speech}   & (47) \textbf{transcript}    & (62) \textbf{audio description}   \\
        (3) customization            & (18) \textbf{header}     & (33) eyestrain                 & (48) default language       & (63) container                     \\
        (4) font size                & (19) overlap             & (34) strain                    & (49) older device           & (64) distinguishable               \\ 
        (5) volume                   & (20) pause button        & (35) background image          & (50) \textbf{visual cue}    & (65) input type                    \\
        (6) \textbf{cannot see}      & (21) \textbf{flicker}    & (36) \textbf{screen reader}    & (51) grouped                & (66) keyboard language             \\
        (7) \textbf{accessibility}   & (22) spacing             & (37) change language           & (52) seizures               & (67) page refresh                  \\
        (8) \textbf{readable}        & (23) migraine            & (38) small widget              & (53) select language        & (68) page title                    \\ 
        (9) change font              & (24) input method        & (39) stop button               & (54) understandable         & (69) sign language        \\
        (10) \textbf{hard to see}    & (25) autoplay            & (40) \textbf{impaired}         & (55) vibration feedback     & (70) svg image                     \\
        (11) background color        & (26) metadata            & (41) \textbf{text reflow}      & (56) actionable             & (71) switch device                 \\
        (12) light mode              & (27) too bright          & (42) timeout                   & (57) audio cue              & (72) touch target                   \\ 
        (13) mute                    & (28) haptic              & (43) consistency               & (58) missing label          & (73) adjust size                    \\
        (14) contrast                & (29) scaling             & (44) epilepsy                  & (59) \textbf{navigable}     & (74) adjust colour                  \\
        (15) subtitle                & (30) control key         & (45) assistance                & (60) verbose                                        \\

         \bottomrule
\end{tabular} 
\end{adjustbox}
\end{center}
\end{table*}

%% file: Tables/Bigrams.tex
\begin{table*}[tbp]
\begin{center}
\caption{
A sample of frequently occuring bigrams for the keywords that are strongly correlated to accessibilty review by our model. }

\label{Table:bigram} 
\begin{adjustbox}{width=1.0\textwidth,center}
\begin{tabular}{llllll}

\toprule
\textbf{Bigram} \\
         \midrule 
\textbf{cannot see} &  \textbf{accessibility} & \textbf{readable} & \textbf{hard to see} \\ \hline
        cannot see anything & easily accessible & readable text & very hard to see  \\ 
        cannot see worksheet & more accessible & document reader & too hard to see   \\ 
        cannot see number & great accessibility & easier reading & really hard to see   \\ 
         cannot see status & accessibility suite  & can read & terribly hard to see \\ 
          still cannot see & accessibility screen& cant read & hard to see theme  \\ 
        \hline
        
       \textbf{blind} & \textbf{header} & \textbf{flicker} &  \textbf{voice command}  \\ \hline
        blind user & theme header & screen flicker & voice command search &   \\ 
        color blind & custom header & flicker taskbar & use voice command  & \\ 
        supports blind & size header & flicker background & voice commands works &  \\
        impaired / blind & adjust header & heavy flickering  & simple voice command &   \\ 
         totally blind & transparent header & constant flickering & custom voice command & &  \\
        \hline
        \textbf{text-to-speech} & \textbf{screen reader} & \textbf{impaired} &
         \textbf{text reflow} \\ \hline 
       verbose text-to-speech  & screen reader accessibility & visually impaired  & text reflow  feature \\
       text-to-speech works & accessibility screen reader & vision impairment  & activate text reflow \\
       text-to-speech feature & talkback screenreader & visual impairment &  good text reflow \\
       text-to-speech news & small-screen reader & great for visually impaired& has text reflow\\
         \hline 
         \textbf{transcript} & \textbf{visual cue} & \textbf{navigable} & \textbf{audio description}  \\ \hline
           transcript title & no visual cue &  navigable bar & turns on audio description\\
           recording / transcription& some visual cue & navigable button & & \\
           zooming and transcript & provide a visual cue & navigable app & &\\
           transcription not found & &  easily navigation & \\  
         \bottomrule
\end{tabular} 
\end{adjustbox}
\end{center}
\end{table*}

%% file: Tables/Misclassification.tex
\begin{table*}
\centering
\caption{
Examples of the misclassification case of our BDTs-model.}
\label{Table:misclassification}
\begin{adjustbox}{width=1.0\textwidth,center}
\begin{tabular}{@{}llrrrl@{}}
\toprule
\multicolumn{1}{c}{\textbf{\begin{tabular}[c]{@{}c@{}}Type\end{tabular}}} & \multicolumn{1}{c}{\textbf{\begin{tabular}[c]{@{}c@{}}Example\end{tabular}}}  \\ \midrule
\multirow{4}{*}{False Positive}  &  \textit{``Simple and easy to use''} \\
& \\
& \begin{tabular}[c]{@{}l@{}} 
\textit{``This app works well - especially ``lucid dream'' - i still remember my dream last week.} 
\\\textit{Amazing!  But i dont like the side effects - like headache and other emotional thing.''} \end{tabular}\\
\midrule
\multirow{7}{*}{False Negative} &  \textit{``Beautiful Functioning Reader''} \\
& \\
& \begin{tabular}[c]{@{}l@{}} 
\textit{``Thank you for all your hard work in making this app for us to use. And to offer it to us for free} \\\textit{is amazing. I use this app everyday, I got all my friends and family using it too. Thank you so} \\\textit{much! I can only think of one thing that could make this app better, if you could add caller ID} \\\textit{with name, and make it so users could turn it on or off, this would be great. Even without that,} \\\textit{this app is great.''} \end{tabular}\\ \bottomrule
\end{tabular}%
\end{adjustbox}
\vspace{-0.4cm}
\end{table*}

%% file: Tables/DifferentApproachComparison.tex
\begin{table*}[h!]
\begin{center}
\caption{Comparison in approaches used to the baselines in our study.}

\label{Table:Approaches comparision}
\begin{adjustbox}{width=1.0 \textwidth,center}
\begin{tabular}{l|ccc|ccc|ccc}
\toprule
\multirow{2}{*}{\textbf{}}  & \multicolumn{3}{c|}{\textbf{Our approach}}                                                                                & \multicolumn{3}{c|}{\textbf{Keyword-based}}                                                                               & \multicolumn{3}{c}{\textbf{Random classifier}}                                                                           \\ 
& \multicolumn{1}{l}{\textbf{Precision}} & \multicolumn{1}{l}{\textbf{Recall}} & \multicolumn{1}{l}{\textbf{F1}} & 
\multicolumn{1}{l}{\textbf{Precision}} & \multicolumn{1}{l}{\textbf{Recall}} & \multicolumn{1}{l}{\textbf{F1}} & \multicolumn{1}{l}{\textbf{Precision}} & \multicolumn{1}{l}{\textbf{Recall}} & \multicolumn{1}{l}{\textbf{F1}}  \\

\midrule


\textbf{Classification} & 0.898 & 0.916 & 0.907         & 0.996 & 0.405 & 0.576             & 0.012 & 0.500 & 0.023 \\ 

\midrule
\textbf{Improvement} & -- & -- & --                     & 0.901 x & 2.261 x & 1.574 x       & 74.833 x & 1.832 x & 39.434 x   \\

\bottomrule
\end{tabular}
\end{adjustbox}
\end{center}
\end{table*}

%% file: Charts/TrainingDataSize.tex
\begin{figure*}[h]  
  \begin{tikzpicture}
   \begin{scope}[scale=1.3]
  
     \begin{axis}[
       width=1.1\columnwidth,
       height=0.7\columnwidth,
       axis x line=bottom,
       xmin=100,
       xmax=5500,
       xlabel={Number of reviews used in the training dataset},
       xlabel near ticks,
       xticklabel style={/pgf/number format/1000 sep=},
       axis y line=left,
       ymin=0,
       ymax=100,
       scaled x ticks=false,
       ylabel={F1-measure (\%)},
       ylabel near ticks, 
       legend style={
         at={(0.5,-0.23)},
            anchor=north,
    	legend columns=2,
            /tikz/every even column/.append style={column sep=0.5cm}
        },
     ]
\addplot[color=blue,mark=triangle*] coordinates {

(100, 53)
(200, 68)
(300, 69)
(400, 73)
(500, 76)
(600, 78 )
(700, 79)
(800, 80 )
(900, 81)
(1000, 82)
(1100, 82)
(1200, 83)
(1300, 84)
(1400, 85)
(1500, 85)
(1600, 86)
(1700, 85)
(1800, 86)
(1900,85 )
(2000, 86)
(2100, 86)
(2200, 86)
(2300, 87)
(2400, 86)
(2500, 87 )
(2600, 86)
(2700, 87)
(2800, 87 )
(2900, 88)
(3000,88 )
(3100,88 )
(3200,88 )
(3300,88 )
(3400,88 )
(3500, 88)
(3600, 88)
(3700, 89)
(3800,88 )
(3900,88 )
(4000,88 )
(4100,88 )
(4200, 89)
(4300, 89)
(4400, 88)
(4500, 89)
(4600, 89)
(4700, 89)
(4800, 89)
(4900, 89)
(5000, 89)
(5100,89 )
(5200, 89)
(5300, 89)
(5326, 89)
       };
     \end{axis}
      \end{scope}
   \end{tikzpicture}
\Description[Line chart presenting the F1-measure of the training dataset]{number of reviews when the dataset incrementally added reviews to the training dataset and evaluated the performance of the classification.}
\caption{F1-measure achieved by incrementally adding training data size for binary classification.} 
\label{Chart:TrainingSizeImpact}
\end{figure*}
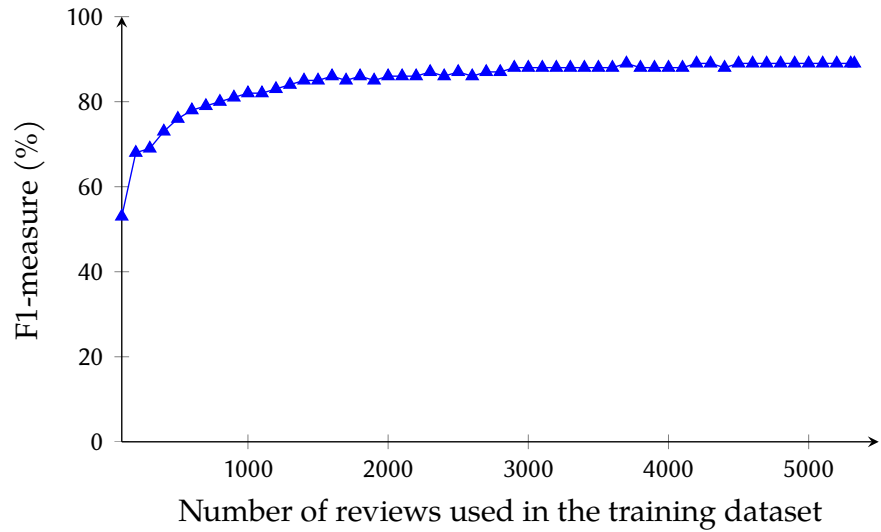